\documentclass[conference]{IEEEtran}
\usepackage{graphicx}
\usepackage{color}
\usepackage{subfigure}
\usepackage{fancyhdr}
\begin{document}

\title{Stochastic Automata Network for Performance Evaluation of Heterogeneous SoC Communication}
\author{\authorblockN{Ulhas Deshmukh}
\authorblockA{Lectrurer in ECE, Govt. Polytechnic, Dhule, India \& \\ Research Scholar MNIT, Jaipur, India\\ 
Email: deshmukhur@gmail.com}
\and
\authorblockN{Vineet Sahula, {\it Senior Member, IEEE}}
\authorblockA{Professor, Deptt. of Electronics \& Comm. Engg.\\Malaviya National Institute of Technology, Jaipur, India\\
Email: sahula@ieee.org}
}
\maketitle
\begin{abstract}
To meet ever increasing demand for performance of emerging System-on-Chip (SoC) applications, designer employ techniques for concurrent communication between components. Hence communication architecture becomes complex and major performance bottleneck. An early performance evaluation of communication architecture is the key to reduce design time, time-to-market and consequently cost of the system. Moreover, it helps to optimize system performance by selecting appropriate communication architecture. However, performance model of concurrent communication is complex to describe and hard to solve. 
In this paper, we propose methodology for performance evaluation of bus based communication architectures, modeling for which is based on modular Stochastic Automata Network (SAN). We employ Generalized Semi Markov Process (GSMP) model for each module of the SAN that emulates dynamic behavior of a Processing Element (PE) of an SoC architecture. 
The proposed modeling approach provides an early estimation of performance parameters viz. memory bandwidth, average queue length at memory and average waiting time seen by a processing element; while we provide parameters viz. number of processing elements, the mean computation time of processing elements and the first and second moments of connection time between processing elements and memories, as input to the model.   

\end{abstract}

\section{Introduction}
\label{intro}

Modern-day System-on-Chip (SoC) platforms use a large number of embedded processors and application specific hardware components \cite{itrs}. An integration of these heterogeneous components into a single chip makes communication among them critical. Besides, these components are pre-verified and optimized. Hence, communication architecture emerges as a key performance determining component of these multiprocessor SoC (MP-SoC) platforms. 
Furthermore, availability of several commercial communication architectures such as, AMBA, CoreConnect and their customization facilitate the designer with variety of design alternatives. Therefore, system level performance estimation is essential for selection of optimum communication architecture from a wide design space at an early stage of design cycle.

System-on-Chip applications use different types of communication architectures viz. bus-based, Network-on-Chip (NoC) based, hybrid bus-NoC architecture and crossbar architecture. Bus based architectures can be further classified as dedicated buses, single shared bus and network of shared buses. In SoCs and embedded applications, bus based architectures are popular because these are simple, consume less power and area. Moreover, performance of bus based architectures not only suffices for low end and high volume applications but also results in cheaper design. This has been motivation for our efforts for estimating performance of bus based communication architectures at the system level.
      	     	     	     	     	     	
In this paper, we propose system level performance estimation of bus based communication architectures based on Stochastic Automata Network (SAN). Mainly, we focus on formulation of    SAN model for a Single Shared Bus (SSB) architecture and its extension for Hierarchical Bus Bridge (HBB) architecture. The approach has been proposed as an extension of GSMP based performance model of these architectures  \cite{selfuk}. In Section \ref{s2}, we present basic concepts and terminology of SAN, related work and our contribution. In Section \ref{s3}, we propose the SAN framework of a SSB architecture for performance estimation. Section \ref{s4} contains enhancement of the SAN formulation for HHB architecture. We present the results in Section \ref{s5}. We conclude in Section \ref{s6}.

\section{Background}
\label{s2}

\subsection{Stochastic Automata Network: an overview}
A stochastic automata network consist of a number of modules or stochastic automata. A module is modeled by a set of states and a set of transitions which determines dynamic behavior of a component of the parallel system. The state of one module is called {\it local state}, while {\it global} or {\it system state} is the collection of local states of all modules. In short, the SAN model is modular representation of parallel system. 
The modules of a SAN model interact with each other using {\it local} and {\it synchronizing events}. Local event changes the state of a single component module by triggering local transition. Synchronizing event modifies the states of more than one modules by simultaneous transitions in those modules. Probabilities of local and synchronizing transition can be {\it functional} or {\it non-functional}. In functional transition, transition probability is the function of the states of other modules whereas it is constant in non-functional transition.

For formal description, let us consider a SAN model with N component modules and a set of events E. The $i^{th}$ automaton, $A^{(i)}$ (where $i=1,2,...,N$) with a set of states $S^{(i)}=\{a^{(i)},...,z^{(i)}\}$ having cardinality $n_i$. Local state variable of $A^{(i)}$ is denoted by $x^{(i)}$. Hence, global state of the SAN is the collection of all local states i.e. a vector $\tilde{x}=(x^{(1)}, x^{(2)},...,x^{(N)}$) whereas $ S= S^{(1)}$ x $S^{(2)}$ x ... x $S^{(N)}$ is called the global state space . The details of SAN can be found in \cite{plateau91} and references there in.

\subsection{Related Work}
Work reported in \cite{daveau95}, uses static performance estimation technique for allocation of communication channels. Our previous work  \cite{selfuk}, proposes an analytical performance evaluation of SSB and HBB architectures based on GSMP model.
Analytical approach as in \cite{knudsen98}, estimates communication overhead in the pipelined communication path, which considers an impact of various protocol parameters on data transfer.  Work in \cite{zhu06} proposes simulation based approach based on Operation State Machine for performance estimation of the system. Authors in \cite{lahiri01} have proposed two phase hybrid performance estimation approach which first performs initial co-simulation with abstract communication and then analyzes time inaccurate communication graph by specifying communication architecture. A large body of work dealing SAN formalization is available in \cite{plateau91} \cite{stewart95}. Authors in \cite{nandi01} use SAN model for performance analysis in platform based design.

\subsection{Contribution of the paper}
Main contribution of the paper lies in the proposal for system level performance estimation of a SSB architecture and HBB architecture. The formulation is based on the SAN model of communication architectures. We present high level simulation model of these architectures in the Stateflow component of MATLAB.

Proposed modeling approach provides an early estimation of memory bandwidth (BW), average queue length ($\overline{L}$) and average waiting time ($\overline{W}$) for a SSB architecture; whereas in case of HBB architecture, we estimate local bandwidth ($BW_\ell$), local average queue length ($\overline{L}_\ell$), local average waiting time ($\overline{W}_\ell$),  global memory bandwidth ($BW_g$), global average queue length ($\overline{L}_g$) and global average waiting time ($\overline{W}_g$).
The input parameters to the model are number of Processing Elements (PEs) (N), the mean computation time ($\overline{T}$) and first and second moment of connection time of PEs ($\overline{C}$, $\overline{C^2}$). 
Additional input parameters for HBB architecture are: probability of local and global requests ($X_\ell$ and $X_g$), first and second moment of local and global connection times ($\overline{C}_\ell$, $\overline{{C_\ell}^2}$, $\overline{C}_g$, $\overline{{C_g}^2}$).

\section{SAN based model for SSB architecture}
\label{s3}
\begin{figure}[h]
\begin{center}
\includegraphics[width=4.5cm,height=3cm] {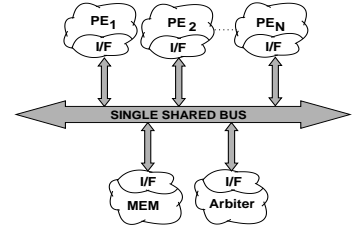}   
\caption{A single shared bus communication architecture.}\label{ssb} 
\end{center}
\end{figure}

In this section, we propose the SAN model of a heterogeneous SSB architecture for evaluating performance metrics. The model has been proposed as an extension of GSMP based performance model of a homogeneous SSB architecture  \cite{selfuk}. 
Two types of abstract communication models are being used in SoC platforms- (i) massage passing communication model and  (ii) shared memory communication model. Our formulation is based on the latter model, in which SoC function involves communication of the PEs with the memories. Figure \ref{ssb} shows synchronous SSB architecture which consists of N heterogeneous processing elements, $PE_1$, $PE_2$,...,$PE_N$ competing for the use of a bus. We assume that a bus arbitration is based on the fixed priorities of PEs. The lowest priority is assigned to $PE_1$ while the highest to $PE_N$. The bus access is assumed to be non-preemptive. Arbiter of N-user one-server type resolves the bus access conflict.  

\begin{figure*}[ht]
\begin{center}\includegraphics[width=5in,height=1.8in]{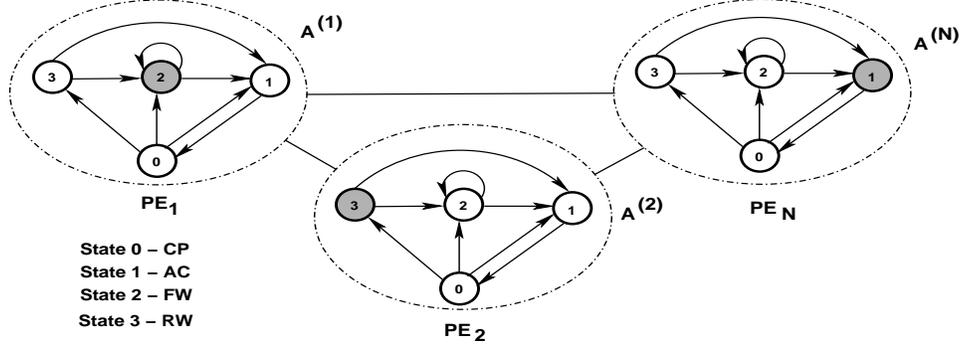} \end{center}
\caption{The SAN model for a heterogeneous SSB communication architecture.} \label{sanssb} 
\end{figure*} 
\subsection{Model formulation}
\begin{figure}[h]
\begin{center}
\scalebox{0.45}{ \input{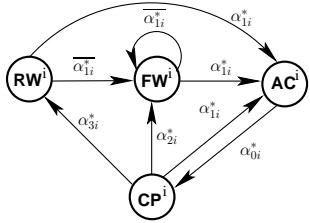}}
\caption{An automaton $A^{(i)}$ representing GSMP model of $PE_i$.} \label{sanpei}
\end{center}
\end{figure}

Stochastic automata network of a heterogeneous SSB architecture is modeled as a collection of interacting modules of PEs. We employ GSMP model \cite{selfuk} for each module which represents dynamic behavior of a PE. We use functional and synchronizing transitions to describe an interactions among these modules. Figure \ref{sanssb} depicts SAN model of a SSB architecture, whereas Fig. \ref{sanpei} shows details of one automaton $A^{(i)}$ that represents GSMP model of $PE_i$. {\it Computing state} labelled as $CP^i$, corresponds to the situation when the $PE_i$ is computing. In {\it Accessing state} $AC^i$, the $PE_i$ accesses MEM.
 In {\it full waiting state} labelled as $FW^i$, the $PE_i$ waits for MEM for full connection time of another PE which is accessing MEM; while in {\it residual waiting state} labelled as $RW^i$, the $PE_i$ waits for MEM for residual connection time of a accessing PE. In each state, model spends random amount of time with mean value $\eta_k$, called mean sojourn time of $k^{th}$ state ($k=CP^i, AC^i, FW^i, RW^i$).

We express state transition probabilities of the SAN model in terms of transition probabilities of GSMP model of a homogeneous SSB architecture \cite{selfuk}. These are explained as follows.
(i) $\alpha_{0i}^*$- a local transition involves only $A^{(i)}$, with constant probability $\alpha_{0i}$. (ii) $\alpha_{1i}^*$- the functional transition which depends on the global state of the system. This transition takes place if all high priorities PEs are in computing states. (iii)  $\alpha_{2i}^*$- a synchronizing transition which synchronizes with event $e_j$ (any $\alpha_1$ transitions of higher priority PEs) with probability $p_e$ and alternate probability 1. (iv) $\alpha_{3i}^*$- a functional transition which takes place if any one of the PEs is in accessing state.
\begin{eqnarray}
\label{tran} 
\alpha_{0i}^*= \alpha_{0i} = 1 \qquad \qquad \qquad \qquad \nonumber \\ 
\alpha_{1i}^*= f(x^j)= \left\{ \begin{array}{ll}
1 & \mbox{if}, \: \: \:  x^j = CP^j, \: j=i+1,...,N \\ 
0 & \mbox{otherwise}  \nonumber \\
\end{array} \right.  \\
\qquad \alpha_{2i}^* = (e_j,p_e,1), \: j=i+1,...,N   \nonumber \\
\alpha_{3i}^* = f(x^j)= \left\{ \begin{array}{ll}
1 & \mbox{if}, \: \: \:  x^j = AC^j, \: j=1,2,...,N \\ 
0 & \mbox{otherwise}  \nonumber \\
\end{array}  \right. \\
\overline{\alpha_{1i}^*} = 1- \alpha_{1i}^*  \qquad \qquad \qquad \qquad \nonumber
\end{eqnarray}
Performance parameters of the $i^{th}$ PE are computed from steady state probabilities  \cite{selfuk}   viz. $BW_i= P_{AC}^i$, $PU_i=P_{AC}^i+P_{CP}^i$, $\overline{L}_i= (P_{FW}^i+P_{RW}^i)$ and $\overline{W}_i=(\eta_{FW}^i \alpha_{2i}^*  + \eta_{RW}^i  \alpha_{3i}^*)/\alpha_{1i}^*$ (where, $P_k$ is steady state probability of the $k^{th}$ state).

\section{SAN based model for HBB architecture}
\label{s4}
\begin{figure}[ht]
\begin{center}\includegraphics[scale=0.9]{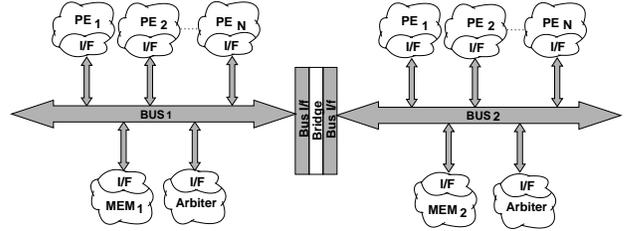} \end{center}
\caption{Hierarchical bus bridge communication architecture.}  \label{f1} 
\end{figure}

In this section, we extend SAN modeling approach for HBB architecture. HBB architecture is composed of two shared buses $BUS_1$ and $BUS_2$, and connected by a bus bridge as shown in Fig. \ref{f1}. Here, N number of PEs on each bus, compete to access shared memories $MEM_1$ or $MEM_2$. At the bridge level communications on two buses are concurrent whereas at bus level behavior of PEs are concurrent.
For simplicity, let us consider a scenario when a PE mapped to $BUS_1$  generates either a local request to access $MEM_1$ or global request to access  $MEM_2$. With reference to this PE, parameters of $MEM_1$ and $MEM_2$ are referred to as local and global parameters, respectively. Let $X_\ell$ be the probability of local request, implying only $BUS_1$ would be used to access $MEM_1$, and arbitration of $BUS_1$ is sufficient. Whereas  $X_g$ be the probability of global request where both $BUS_1$ and $BUS_2$ would be used to access $MEM_2$, and two stage arbitration of $BUS_1$ and $BUS_2$ is essential. 
         
\subsection{Model formulation}
We propose two level SAN model for HBB architecture. At bridge level the SAN consist of two automata correspond to $BUS_1$ and $BUS_2$ and are similar to the Fig. \ref{sanssb}. At bus level, each module is composed of automata of PEs. At bridge level two automata of buses interact with each other while at bus level interaction among automata of PEs is modeled.   

\begin{figure}[h]
\begin{center}
\scalebox{0.55} {\input{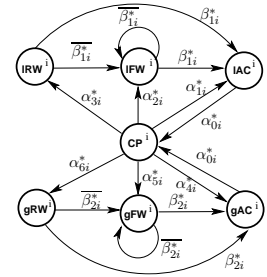}}
\caption{An automaton $A^{(i)}$ of $PE_i$ in HBB architecture.} \label{f2}
\end{center}
\end{figure}

Automata of the $PE_{1i}$ in aforementioned scenario (mapped to $BUS_1$) is depicted in Fig. \ref{f2}.
{\it State} $lAC_i$, {\it state} $lFW_i$ and {\it state} $lRW_i$ correspond to local memory  $MEM_1$ and are similar to the states of automata of a $PE_i$ of SSB architecture (Fig. \ref{sanpei}). Global accessing state labelled as {\it state} $gAC_i$, global full waiting state labelled as {\it state} $gFW_i$ and global residual waiting state labelled as {\it state} $gRW_i$ are analogous states when a PE attempts to access $MEM_2$. Detail discussion of model equations and performance parameters is omitted. 

\section{Results}
\label{s5}
In this section, we present performance evaluation results of SSB and HBB architectures obtained using the proposed modeling approach. We have captured the SAN model of both architectures with fixed arbitration scheme in Stateflow component of MATLAB. Simulation was performed on on P-IV, 1 GB Linux-workstation. In both examples, random computation and communication times of PEs were generated by using MATLAB m-functions with generalized distribution. 

\begin{figure}[ht]
\begin{center}\includegraphics[width=3.4in,height=1.5in]{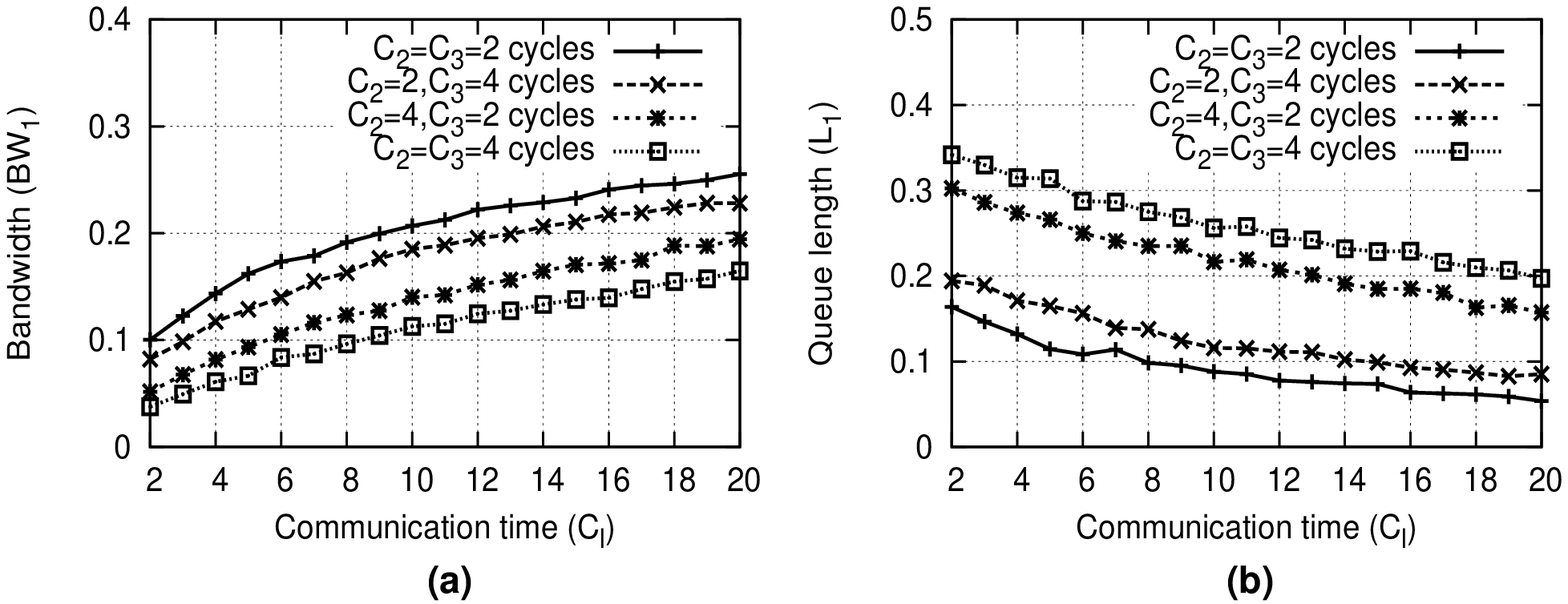} \end{center}
\caption{Variation of (a) $BW_1$ and (b) $\overline{L}_1$, with $C_1$.}  \label{r1} 
\end{figure}

As first example, we have considered a SSB architecture with three PEs- $PE_1$, $PE_2$ and $PE_3$. We assigned the lowest  priority to $PE_1$ and the highest to $PE_3$. We assigned mean values of computation times of PEs as: $\overline{T}_1=\overline{T}_2=\overline{T}_3$= 2 cycles. We varied mean communication time ($\overline{C}_1$) of $PE_1$ with $\overline{C}_2$ and $\overline{C}_3$ as parameters. 
Various performance parameters of the PEs viz. $BW$, $\overline{L}$ and $\overline{W}$ have been estimated. For brevity, we present results of $BW_1$ and $\overline{L}_1$ of $PE_1$, as shown in Fig. \ref{r1}(a) and \ref{r1}(b). 

As observed from the Fig. \ref{r1}(a), bandwidth increases with communication time which is due to increase in mean sojourn time of $AC^1$ state. The Fig. also shows influence of $\overline{C}_2$ and/or $\overline{C}_3$ on $BW_1$. Reduction in bandwidth is observed when we changed $\overline{C}_2$ and/or $\overline{C}_3$ from two to four cycles, since $PE_1$ has to wait more time in waiting states. $PE_1$ received maximum bandwidth (25 \%) when $\overline{C}_2$=$\overline{C}_3$=2 cycles and $\overline{C}_1$=20 cycles; and minimum bandwidth (3 \%) when $\overline{C}_2$ = $\overline{C}_3$ = 4 cycles and $\overline{C}_1$=2 cycles. Figure \ref{r1}(b) reveals converse observations for queue length, $\overline{L}_1$. For higher values of $\overline{C}_2$ and/or $\overline{C}_3$, $PE_2$ and/or $PE_3$ access MEM for more time than $PE_1$. As a consequence $PE_1$ spends more time in waiting states. Hence, higher value of $\overline{L}_1$ is noted for $\overline{C}_2$ = $\overline{C}_3$ = 4 cycles. 

\begin{figure}[ht]
\begin{center}\includegraphics[width=3.4in,height=1.5in]{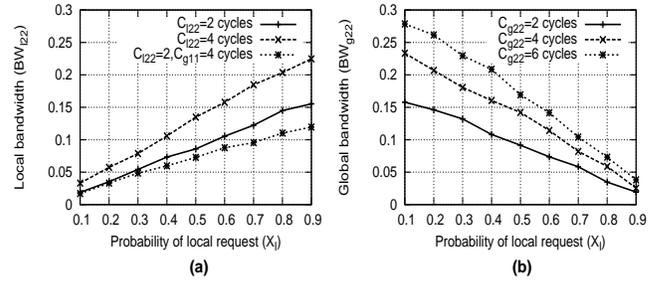} \end{center}
\caption{Effect of $X_{\ell22}$ on (a) $BW_{\ell22}$ and (b) $BW_{g22}$. }  \label{r2} 
\end{figure}

In second example, we have considered a HBB architecture with two PEs mapped to each bus. Processing elements, $PE_{11}$ and $PE_{12}$ are mapped to $BUS_1$; while $PE_{21}$ and $PE_{22}$ are mapped to $BUS_2$. We assigned descending priorities from global requests of $PE_{22}$, $PE_{21}$, $PE_{12}$ and $PE_{11}$; and then local requests in the same order. Various model input parameters are assigned values as follows- $X_{\ell11}$=0.7, $X_{\ell12}$=0.8, $X_{\ell21}$=0.7, $\overline{T}_{11}$=$\overline{T}_{12}$=$\overline{T}_{21}$=$\overline{T}_{22}$= 2 cycles, $\overline{C}_{\ell11}$=$\overline{C}_{\ell12}$=$\overline{C}_{\ell21}$= 2 cycles, and $\overline{C}_{g11}$=$\overline{C}_{g12}$=$\overline{C}_{g21}$= 2 cycles (here, $\ell$ and g denote local and global parameters followed by PE number).
From various evaluated performance parameters of PEs, we present local and global bandwidth ($BW_{\ell22}$,$BW_{g22}$) of $PE_{22}$. We varied $\overline{C}_{\ell22}$  for local bandwidth and $\overline{C}_{g22}$ for global bandwidth. Figure \ref{r2}(a) and \ref{r2}(b) show plot of these parameters with probability of local request, $X_{\ell22}$ with $\overline{C}_{\ell22}$ and  $\overline{C}_{g22}$ as parameters.

We observe that local bandwidth, $BW_{\ell22}$ increases with increase in $X_{\ell22}$ as well as with $\overline{C}_{\ell22}$. At higher values of  $X_{\ell22}$, $BW_{\ell22}$ is more sensitive to $\overline{C}_{\ell22}$. An influence of $\overline{C}_{g11}$ on $BW_{\ell22}$ is clearly noted from the Fig. \ref{r2}(a). Share of local bandwidth declined as we increased $\overline{C}_{g11}$ from two cycles to four cycles. In case of global bandwidth, $BW_{g22}$ gradual decrease is observed with increase in $X_{\ell22}$. At the same value of $X_{\ell22}$, the $PE_{22}$ received  more bandwidth with higher $\overline{C}_{g22}$. Variations in $BW_{g22}$ with $\overline{C}_{g22}$ at higher values of  $X_{\ell22}$ are not significant.

\section{Conclusions}
\label{s6}
This paper presents SAN based modeling approach for system level performance evaluation of SSB and HBB architectures. We have evaluated performance metric viz. bandwidth, queue length and waiting time with communication times of processing elements for SSB architecture. For HBB architecture performance parameters for local and global memories are evaluated with local requesting probabilities. Proposed approach provides an early estimation of performance metrics that can help the designer to select the appropriate communication architecture for SoC and embedded applications.\\
\subsubsection*{\bf Acknowledgments}
\small{
\noindent We gratefully acknowledge the financial support provided by the Department of IT, Ministry of Communication  \& IT, Govt. of India under SMDP-VLSI-II project.}
\bibliographystyle{IEEEtran}

\bibliography{nor}

\end{document}